\documentclass[12pt]{article}
\usepackage{graphicx}

\def\pbnr{}
\def\speaker{Jingzhi Zhang}
\def\onbehalfof{the BESIII collaboration}
\def\title{Experimental overview of charmonium spectroscopy}
\def\affiliation{Institute of High Energy Physics\\
Chinese Accademy of Sciences, Beijing 100049, China}
\def\support{The workshop was supported by the University of Manchester, IPPP, STFC, and IOP}
\newcommand{\piz}{\pi^0}

\newcommand{\etac}{\eta_c}
\newcommand{\thehc}{h_c}

\newcommand{\psp}{\psi^{\prime}}

\newcommand{\chico}{\chi_{c1}}
\newcommand{\chict}{\chi_{c2}}

\newcommand{\jpsi}{J/\psi}

\newcommand{\pp}{\pi^+\pi^-}
\newcommand{\kk}{K^+K^-}
\newcommand{\ks}{K_S}
\newcommand{\kkp}{K\bar{K}\pi}
\newcommand{\kskp}{K_SK^+\pi^-}
\newcommand{\kkpiz}{\kk\piz}
\newcommand{\kskppp}{K_SK^+\pp\pi^-}
\newcommand{\kkpppiz}{\kk\pp\piz}

\newcommand{\pppppp}{3(\pp)}

\newcommand{\etapp}{\eta\pp}

\newcommand{\mev}{\mathrm{MeV}}

\newcommand{\mevcc}{\mathrm{MeV}/c^2}

\newcommand\pubnumber{\pbnr}
\newcommand\pubdate{\today}
\def\Title#1{\begin{center} {\Large #1 } \end{center}}
\def\Author#1{\begin{center}{ \sc #1} \end{center}}

\newcommand{\OnBehalf}[1]{\sbox0{#1}\ifdim\wd0=0pt
        {}% if #1 is empty
	\else
	{\\on behalf of #1}% if #1 is not empty
	\fi}
\newcommand{\SupportedBy}[1]{\sbox0{#1}\ifdim\wd0=0pt
        {}% if #1 is empty
	\else
	{\footnote{#1}}% if #1 is not empty
	\fi}
\def\Address#1{\begin{center}{ \it #1} \end{center}}

\newcommand\pubblock{\includegraphics[width=5cm]{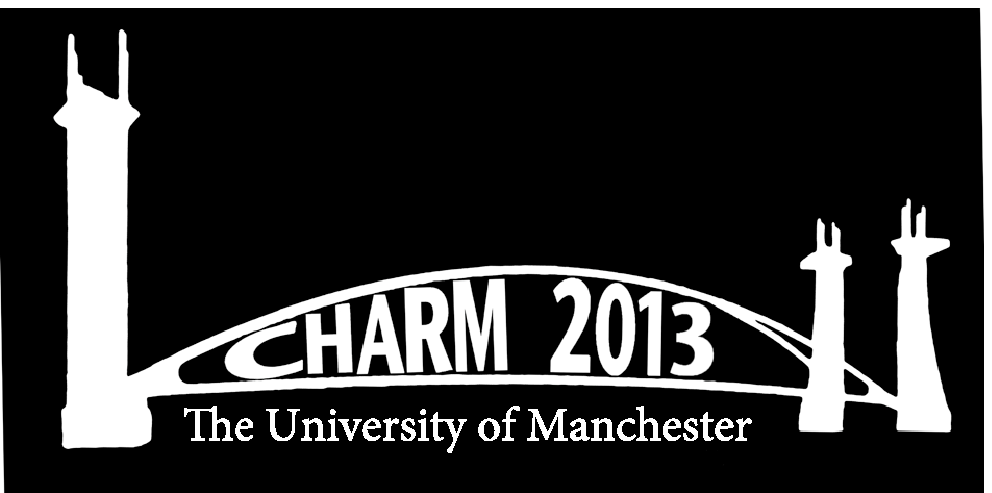}\hfill{\begin{tabular}{l} \pubnumber\\
         \pubdate  \end{tabular}}}
\newenvironment{Abstract}{\begin{quotation}  }{\end{quotation}}
\newenvironment{Presented}{\begin{quotation} \begin{center} 
             PRESENTED AT\end{center}\bigskip 
      \begin{center}\begin{large}}{\end{large}\end{center} \end{quotation}}
\def\Acknowledgements{\bigskip  \bigskip \begin{center} \begin{large}
             \bf ACKNOWLEDGEMENTS \end{large}\end{center}}
\def\venue{The 6$^{th}$ International Workshop on Charm Physics\\
(CHARM 2013)\\
Manchester, UK,  31 August -- 4 September, 2013}
\def\beq{\begin{equation}}
\def\eeq#1{\label{#1}\end{equation}}
\def\eeqn{\end{equation}}

\def\beqa{\begin{eqnarray}}
\def\eeqa#1{\label{#1}\end{eqnarray}}
\def\eeqan{\end{eqnarray}}

\let\bar=\overbar

\def\Dslash{\not{\hbox{\kern-4pt $D$}}}
\def\dslash{\not{\hbox{\kern-2pt $\del$}}}

\def\msb{{\bar{\ssstyle M \kern -1pt S}}}

\begin{document}
\begin{titlepage}
\pubblock

\vfill
\Title{\title}
\vfill
\Author{\speaker\SupportedBy{\support}\OnBehalf{\onbehalfof}}
\Address{\affiliation}
\vfill
\begin{Abstract}
%%%%%%%%%%%%%%%%%%%%%%%%%%%%%%%%%%%%%%%%%%%%%%%%%%%%%%%%%%%%%%%%%%%%%%%%%%%
% YOUR ABSTRACT GOES HERE
%%%%%%%%%%%%%%%%%%%%%%%%%%%%%%%%%%%%%%%%%%%%%%%%%%%%%%%%%%%%%%%%%%%%%%%%%%%
In this talk, I review the recent experimental results on charmonium
spectroscopy from BESIII, Belle, BaBar and CLEOc. 
Below the open-charm threshold, the masses and widths of spin-singlet
states $\eta_c$, $\eta_c(2S)$, $h_c$ are measured with high precision.
Above the threshold, $\chi_{c2}(2P)$ is identified in
the process $\gamma\gamma \to\chi_{c2}(2P) \to \gamma D \bar{D}$; 
Evidence of $X(3823)$ is found in the $M(\chi_{c1}\gamma)$
invariant-mass distribution for $B\to \gamma\chi_{c1}K$ decays,
the measured properties are consistent with the missing
$\psi_2(1^{2}D_2)$ state.

\end{Abstract}
\vfill
\begin{Presented}
\venue
\end{Presented}
\vfill
\end{titlepage}
\def\thefootnote{\fnsymbol{footnote}}
\setcounter{footnote}{0}
%
%%%%%%%%%%%%%%%%%%%%%%%%%%%%%%%%%%%%%%%%%%%%%%%%%%%%%%%%%%%%%%%%%%%%%%%%%%%
%  WHAT FOLLOWS IS YOUR TEXT
%%%%%%%%%%%%%%%%%%%%%%%%%%%%%%%%%%%%%%%%%%%%%%%%%%%%%%%%%%%%%%%%%%%%%%%%%%%
\section{Introduction}
Charmonium spectroscopy is an ideal place for studying the dynamics of
quantum chromodynamics(QCD) in the interplay of perturbative and
non-perturbative QCD regime. The non-relativistic model including
color Coulomb, linear scalar potential, spin-spin, and spin-orbit
interaction, has made great successes in the description of the
charmonium spectrum \cite{1974yr}.
A relativised version was developed by Godfrey and Isgur
\cite{Godfrey:1985xj} by taking into account relativistic correction
and other effects.
States with $J^{PC} = 1^{--}$ can be produced directly through
electron positron annihilation, and present themselves as enhancements
in the total hadronic cross section. 
Apart from the widely studied $J/\psi$ and $\psi(2S)$ states,
the vectors $\psi(4040)$, $\psi(4415)$, and $\psi(3770)$  have also been found
\cite{Siegrist:1976br,Rapidis:1977cv,Bacino:1977uh,Brandelik:1978ei,BES_R_Value},
assigned as $\psi$($3^3S_1$), $\psi$($4^3S_1$),
$\psi$($1^3D_1$) and $\psi$($2^3D_1$), respectively.
All predicted charmonium states below the $DD$ threshold have been
observed in experiment. Among them, the spin-singlet states $\eta_c$,
$\eta_c(2S)$, $h_c$ are poorly measured. 
While above the threshold, experimental information on the charmonium
states is rather limited.

%%%%%%%%%%%%%%%%%%%%%%%%%
\section{The $\etac(1S)$}
%%%%%%%%%%%%%%%%%%%%%%%%%
The mass and width of the lowest lying charmonium state, the $\etac$ ($1^1S_0$),
continue to have large uncertainties when compared to those of other
charmonium states~\cite{PDG}. Early measurements of the properties of
the $\etac$ using $\jpsi$ radiative
transitions~\cite{Baltrusaitis:1985mr,Bai:2003et} found a mass and
width of $2978~\mevcc$ and $10~\mev$, respectively. However, recent
experiments, including photon-photon fusion and $B$ decays, have
reported a significantly higher mass and a much wider
width~\cite{Aubert:2003pt, Asner:2003wv,  Uehara:2007vb, 2011dy}. 
The most recent study by the CLEOc experiment,
using both $\psp \to \gamma\etac$ and $\jpsi\to \gamma\etac$, pointed
out a distortion of the $\etac$ line shape in $\psp$
decays~\cite{recent:2008fb}. 
CLEOc attributed the $\etac$ line-shape distortion to the energy
dependence of the ``hindered'' $M1$ transition matrix element.

\begin{figure}
\begin{center}
  \includegraphics[width=150mm]{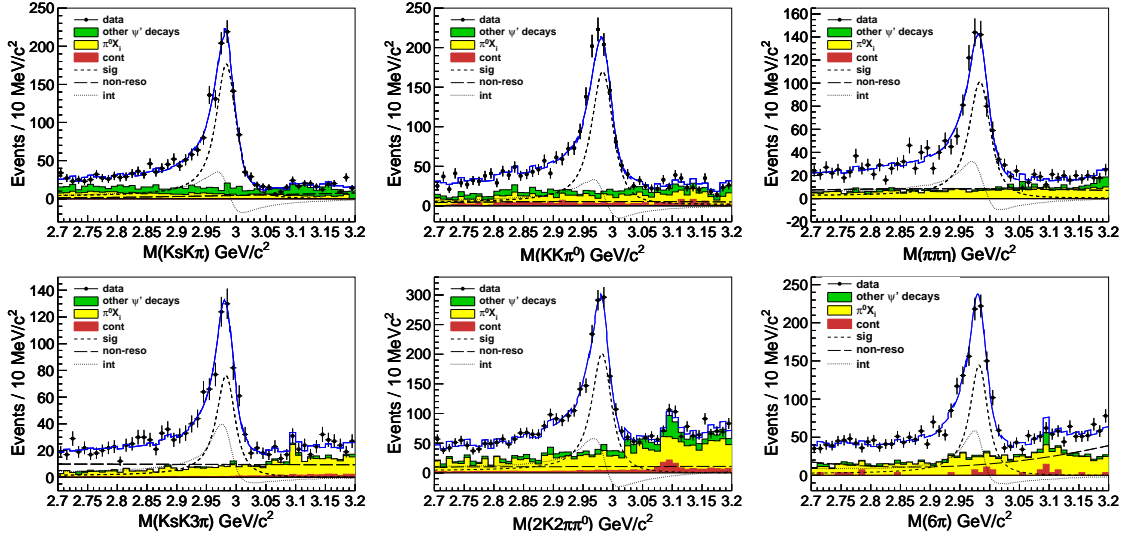}
  \caption{The $M(X_i)$ invariant-mass distributions for the $\etac$
  decays to  $\kskp$, $\kkpiz$, $\etapp$, $\kskppp$, $\kkpppiz$ and $\pppppp$,
  respectively, with the fit results superimposed. 
  Points are data and the solid lines are the total fit results. 
  %Signals are shown as
  %short-dashed lines; the non-resonant components as long-dashed
  %lines; and the interference between them as dotted lines.
  %Shaded histograms are (in red/yellow/green) for (continuum/$\pi^0
  %X_i$/other $\psp$ decays) backgrounds. 
  The continuum backgrounds for $\kskp$ and $\etapp$ decays are negligible.}
  \label{fig:metac}
\end{center}
\end{figure}

At BESIII, the $\etac$'s are produced through $\psp \to \gamma\etac$,
and the $\etac$ mass and width are determined \cite{BESIII:2011ab} by fits to the
invariant-mass spectra of exclusive $\etac$ decay modes. 
Six modes are used to reconstruct the $\etac$: $\kskp$, $\kkpiz$, $\etapp$,
$\kskppp$, $\kkpppiz$, and $\pppppp$, where the $\ks$ is reconstructed
in $\pp$, and the $\eta$ and $\piz$ in $\gamma\gamma$ decays.
Figure~\ref{fig:metac} shows the $\etac$ invariant-mass
distributions for selected $\etac$ candidates, together with the
estimated $\piz X_i$ backgrounds ($X_i$ represents the $\etac$ final
states under study), the continuum backgrounds normalized by
luminosity, and other $\psp$ decay backgrounds estimated from the
inclusive MC sample.
A clear $\etac$ signal is evident in every decay mode. The $\etac$
signal has an obviously asymmetric shape that suggests possible
interference with a non-resonant $\gamma X_i$ amplitude.
%In this analysis, a 100\% of the non-resonant amplitude interferes 
%with the $\etac$ is assumed.
%
The fitted relative phases between the signal and the non-resonant component
from each mode are consistent within $3\sigma$,
which may suggest a common phase in all the modes under study. A fit
with a common phase (i.e. the phases are constrained to be the same)
describes the data well.
The results on the $\etac$ mass and width  are
$M    = 2984.3\pm 0.6 ({\rm stat.}) \pm 0.6({\rm syst.})~\mathrm{MeV}/c^2$ and
$\Gamma = 32.0\pm 1.2 ({\rm stat.}) \pm 1.0({\rm syst.})~
\mathrm{MeV}$, respectively.

The measurements of $\etac$ mass and width are
consistent with those from two photon-photon fusion, $\psp$
transition, and $B$ decays.
Hyperfine splitting is
$\Delta M(1S) = 112.5\pm 0.8~\mevcc$
which agrees well with recent lattice computations~\cite{arXiv:1211.2253}.

%%%%%%%%%%%%%%%%%%%%%%%%%%%%%%%%%
\section{The $\thehc$ $(1^1P_1)$}
%%%%%%%%%%%%%%%%%%%%%%%%%%%%%%%%%
Information about the spin-dependent interaction of heavy quarks
can be obtained from precise measurement of the $1P$ hyperfine mass
splitting $\Delta M_{hf}\equiv\langle M(1^3P)\rangle- M(1^1P_1)$,
where $\langle
M(1^3P_{J})\rangle=(M(\chi_{c0})+3M(\chi_{c1})+5M(\chi_{c2}))/9=
3525.30\pm0.04$ $\mevcc$ \cite{PDG} is the spin-weighted centroid of
the $^3P_J$ mass and $M(1^1P_1)$ is the mass of the singlet state
$\thehc$. A non-zero hyperfine splitting may give an indication of non-vanishing
spin-spin interactions in charmonium potential models~\cite{swanson}.
CLEOc and BESIII measured \cite{Dobbs:2008ec,Ablikim:2010rc} the
$\thehc$ in the $\piz$ recoil mass distribution for $\psp\to\piz
\thehc$ with and without the subsequent radiative decay
$\thehc\to\gamma\etac$ previously.

\begin{figure}[bpt]
\begin{center}
  \includegraphics[height=75mm]{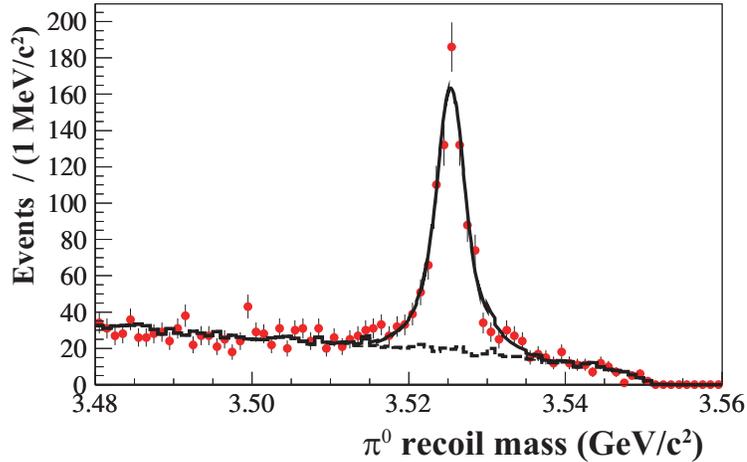}
  \caption{The $\pi^0$ recoil mass spectrum in
    $\psi(3686)\to\pi^0{h}_c, {h}_c\to\gamma\eta_c$, $\eta_c \to X_i$
    summed over the 16 final states $X_i$.  The dots with error bars
    represent the $\pi^0$ recoil mass spectrum in data. The solid line
    shows the total fit function and the dashed line is the background
    component of the fit.}
  \label{hc-FitData}
\end{center}
\end{figure}

In order to reduce background and improve the precision,
BESIII uses 16 exclusive hadronic $\etac$ decay modes to reconstruct
$\thehc\to \gamma\etac$ \cite{Ablikim:2012ur}, where
the 16 hadronic final states include
$p\bar{p}$, $2(\pi^+ \pi^-)$, $2(K^+ K^-)$, $K^+ K^- \pi^+ \pi^-$, $p
\bar{p} \pi^+ \pi^-$, $3(\pi^+ \pi^-)$, $K^+ K^- 2(\pi^+ \pi^-)$,
$K^+ K^- \pi^0$, $p \bar{p}\pi^0$, $\ks K^\pm \pi^\mp$, $\ks K^\pm
\pi^\mp \pi^\pm \pi^\mp$, $\pi^+ \pi^- \eta$, $K^+ K^- \eta$,
$2(\pi^+ \pi^-) \eta$, $\pi^+ \pi^- \pi^0 \pi^0$, and $2(\pi^+
\pi^-) \pi^0 \pi^0$.
By doing so, the ratio of signal to background can be improved significantly. 
A simultaneous fit to the $\piz$ recoil mass distributions of the 16
decay modes is performed to extract signal, the sum of all the
decay modes is shown in Fig.~\ref{hc-FitData}.
From 106 M $\psp$, $835\pm 35$ signal events are obtained. The
measured $\thehc$ mass and width are 
$M  =3525.31\pm0.11\pm0.14~\mevcc$, 
$\Gamma =0.70\pm0.28\pm0.22~\mev$, 
and mass splitting $\Delta
M(1^1P_1)=-0.01\pm0.11~\pm0.15~\,\rm{MeV/c^2}$
which is consistent with the lowest-order expectation that the $1P$ hyperfine splitting is zero.
Where the first errors are statistical and second are systematic.
The results are in agreement with the inclusive analysis results
\cite{Dobbs:2008ec,Ablikim:2010rc}.

\begin{figure}[bpt]
\begin{center}
  \includegraphics[width=75mm]{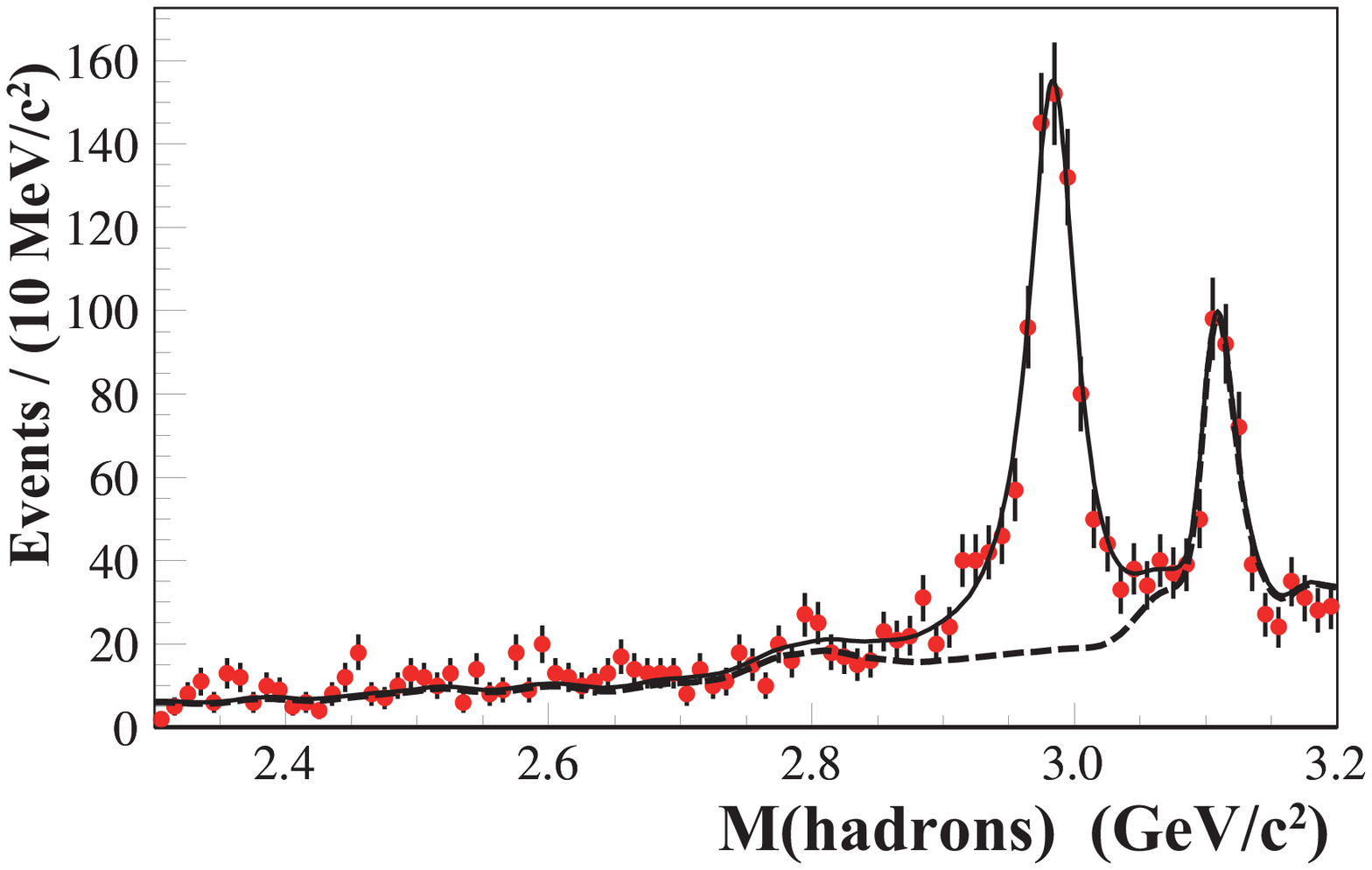}
  \includegraphics[width=75mm]{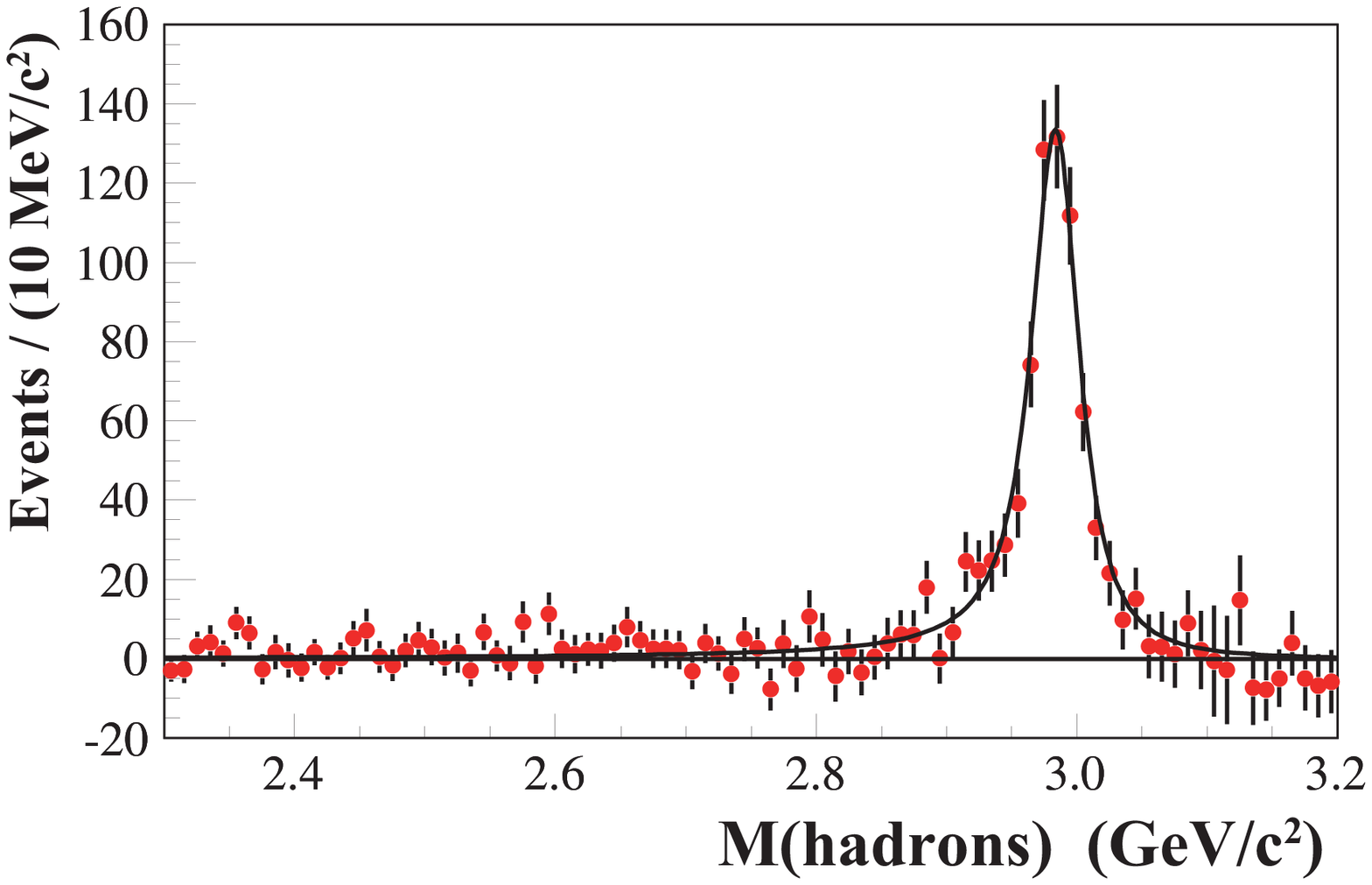}
  \caption{
Left: the hadronic mass spectrum in $\psp\to\pi^0{h}_c, {h}_c\to\gamma\eta_c$,
$\eta_c \to X_i$ summed over the 16 final states $X_i$.  
Right: the background-subtracted hadronic mass spectrum with the
signal shape overlaid.
The dots with error bars
represent the hadronic mass spectrum in data.  The solid line shows
the total fit function and the dashed line is the background component
of the fit. 
}
  \label{fig:etac_fitting}
\end{center}
\end{figure}

%%%%%% the lineshape %%%%%%%
Figure~\ref{fig:etac_fitting} shows the hadronic mass spectrum.
We notice the $\eta_c$ line shape in the $E1$ transition
$h_c\to\gamma\eta_c$ is not as distorted as in the $\psp \to \gamma \eta_c$
decays (as seen in Fig.~\ref{fig:metac}).
The branching ratio of $h_c\to\gamma\eta_c$ is about $50\%$(branching
ratio of $M1$ transition $\psp\to \gamma\eta_c$ is about $0.3\%$), 
non-resonant interfering backgrounds to the dominant transition is
small. With the larger $\psp$ data sample and the advantage of
negligible interference effects, we expect that $h_c\to\gamma\eta_c$
will provide the most reliable determinations of the $\eta_c$
resonance parameters in the future.

%%%%%%%%%%%%%%%%%%%%%%%%%
\section{The $\etac(2S)$}
%%%%%%%%%%%%%%%%%%%%%%%%%
The radially excited $n=2$ spin-singlet $S$-wave state, the $\etac(2S)$ meson,
was not well established until the Belle collaboration
found the $\etac(2S)$ signal at $3654\pm 6({\rm stat.})\pm
8({\rm syst.})$ $\mevcc$ in the $\kskp$ invariant-mass distribution in a
sample of exclusive $\etac(2S)\to \kskp$ decays~\cite{Choi:2002na}. 
Since then measurements of $\etac(2S)$ in photon-photon fusion to
$K\bar{K}\pi$ final state have been
reported~\cite{Aubert:2003pt,Asner:2003wv,Nakazawa:2008zz}, as well as
in double charmonium production \cite{Aubert:2005tj,Abe:2007jn}.
CLEOc searched for $\etac(2S)$ in the radiative decay
$\psp\to\gamma\etac(2S)$, found no clear signals in its sample of 25M
$\psp$~\cite{:2009vg}. The challenge of this measurement is the detection of 50 $\mev$ photons.

With 519 fb$^{-1}$, BaBar observed~\cite{delAmoSanchez:2011bt}
$\etac(2S)\to\kskp$ and $\etac(2S)\to \kkpppiz$ produced in
photon-photon fusion for the first time.
They measured the mass and width of $\etac$ and $\etac(2S)$ in $\kskp$
decays, $M(\etac)= 2982.5 \pm 0.4 \pm 1.4~\mevcc$,
$\Gamma(\etac) = 32.1 \pm 1.1 \pm 1.3~\mev$,
$M(\etac(2S))= 3638.5 \pm 1.5 \pm 0.8~\mevcc$, $\Gamma(\etac(2S)) = 13.4 \pm 4.6 \pm 3.2~\mev$.
%These $\etac(2S)$ results are so far the most precise measurements.

%Belle results for B->K Etac(1S,2S);
Belle updated the analysis of $B^\pm\to K^{\pm}\etac$ and $B^{\pm} \to
K^\pm \etac(2S)$ followed by $\etac$ and $\etac(2S)$ decay to $\kskp$
with 535 million $B\bar{B}$-meson pairs~\cite{2011dy}.
Both decay channels contain the backgrounds from $B^\pm \to K^\pm\kskp$ decays
without intermediate charmonia, which could interfere with the
signal. Belle's analysis took interference into account with no
assumptions on the phase or absolute value of the interference. 
A two dimensional $M(\kskp)$ -- $\cos\theta$ fit was performed to extract signal,
where $\theta$ is the angle between $K$ (from $B$ directly) with
respect to $\ks$ in the rest frame of the $\kskp$.
They obtained the masses and widths of $\eta_c$ and $\eta_c(2S)$.
For the $\eta_c$ meson parameters the model error is negligibly small:
$M(\eta_c)=2985.4\pm1.5({\rm stat.})^{+0.2}_{-2.0}({\rm syst.})$ MeV/$c^2$,
$\Gamma(\eta_c)=35.1\pm 3.1({\rm stat>})^{+1.0}_{-1.6}({\rm syst.})$ MeV/$c^2$.
For the $\eta_c(2S)$ meson the model and statistical uncertainties cannot
be separated:
$M(\eta_c(2S))=3636.1^{+3.9}_{-4.1}({\rm stat.+model})^{+0.5}_{-2.0}({\rm syst.})$ MeV/$c^2$,
$\Gamma(\eta_c(2S))=6.6^{+8.4}_{-5.1}({\rm stat.+model})^{+2.6}_{-0.9}({\rm syst.})$ 
MeV/$c^2$.

\begin{figure}[bpt]
  \includegraphics[width=80mm]{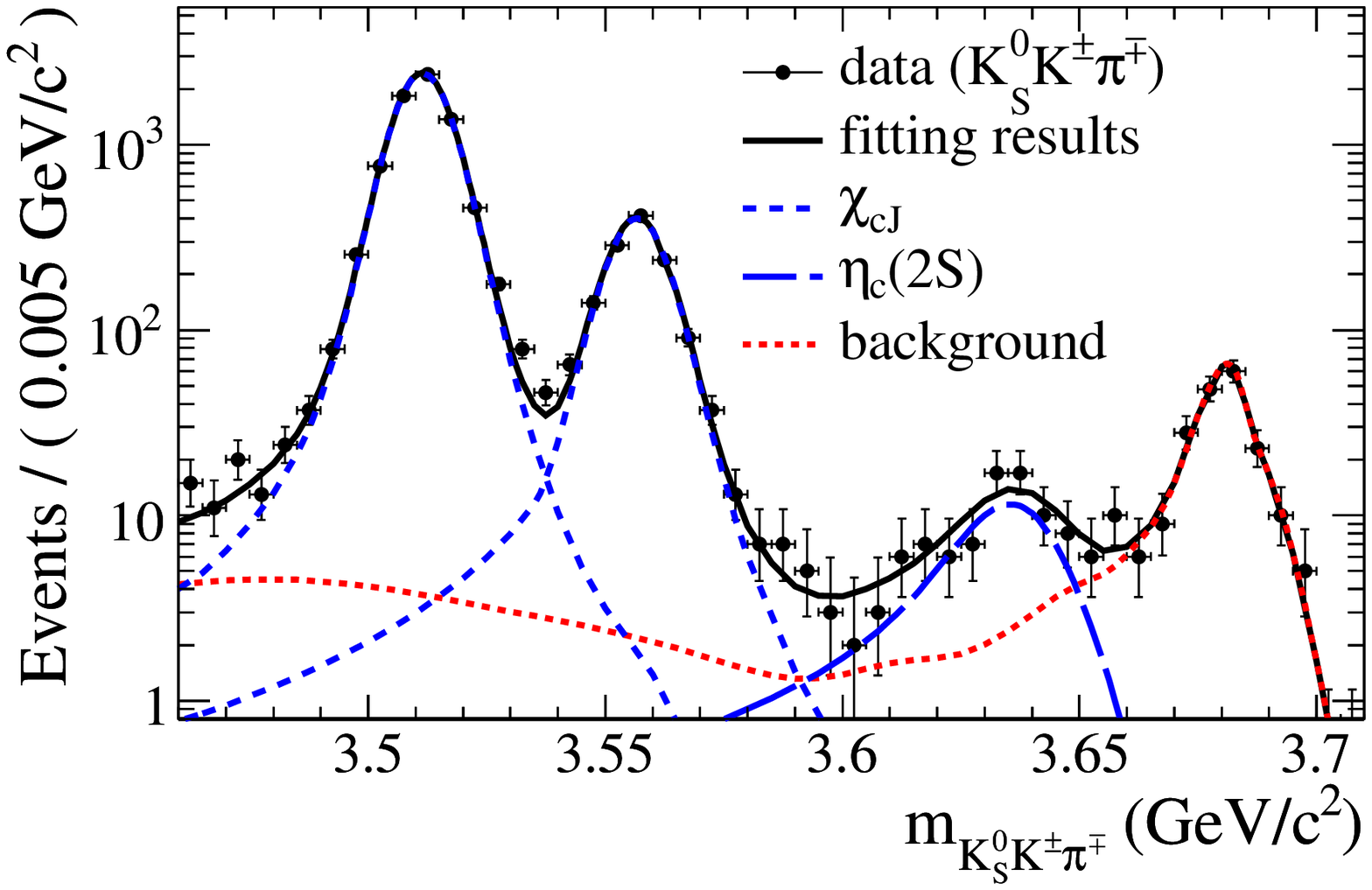}
  \includegraphics[width=80mm]{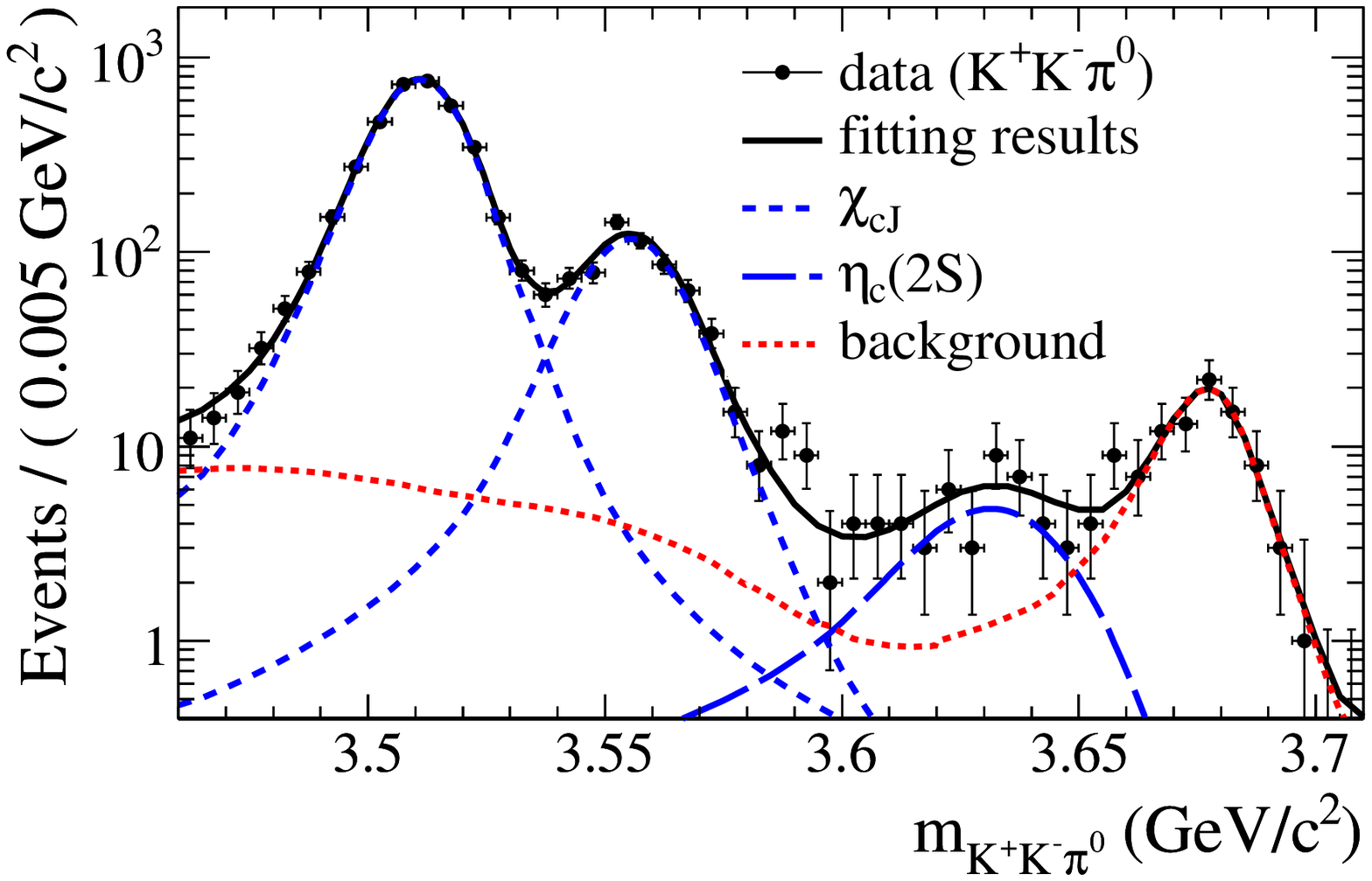}
  \caption{The $\kskp$ invariant mass for selected events for $\psp\to
    \gamma\kskp$ (left) and $KK\pi^0$ (right). Points are data and the
    solid curve is the fit results. Blue long-dashed line is
    signal. Blue dashed lines are $\chico/\chict \to \gamma\kskp$
    events. Red dotted line is for other backgrounds mainly from the
    decays $\psp \to \piz\kskp$, $\kskp$ and ISR/FSR production of
    $\kskp\gamma_{ISR}/\gamma_{FSR}$.}
  \label{fit_etacp}
\end{figure}

% BESIII etac(2S) 
Using 106 million $\psp$ events, BESIII searches \cite{Ablikim:2012sf}
for $\etac(2S)$ in the decay $\psp\to \gamma\etac(2S)$, with
$\etac(2S)\to \kkp$.
Figure \ref{fit_etacp} shows the invariant-mass distributions of
$\kskp$ (left) and $\kkpiz$ (right), where a three-constraints
kinematic fit has been applied (in which the energy of the photon is
allowed to float). The solid curve in Fig.~\ref{fit_etacp} shows
fitting results of an unbinned maximum likelihood fit with four
components: signal, $\chi_{c1}$, $\chi_{c2}$ and other background (coming from $\psp$
decays to $\piz\kkp$, $\kkp$ and ISR/FSR production of $\kkp\gamma_{ISR}/\gamma_{FSR}$).
The fit yields $81\pm14$ signal events for the $\kskp$ channel and $46\pm 11$ for the
$\kkpiz$ channel, and gives the mass $M(\etac(2S)) =3637.6
\pm2.9\pm1.6~\mevcc$ and width $\Gamma(\etac(2S)) = 16.9\pm 6.4\pm4.8~\mev$.
The statistical significance of the signal is more than 11.1 $\sigma$.
Using the detection efficiency determined from MC simulation, the
product branching fraction is obtained 
${\cal B} (\psp\to\gamma\etac(2S)) \times {\cal B}(\etac(2S)\to \kkp)
=(1.30\pm 0.20\pm0.30)\times 10^{-5}$.
Using the result ${\cal B}(\etac(2S)\to \kkp) =(1.9\pm0.4\pm
1.1)\%$ from BaBar~\cite{Aubert:2008kp} gives the branching fraction
${\cal B}(\psp\to\gamma\etac(2S)) =(6.8\pm1.1\pm 4.5)\times 10^{-4}$.
This result is consistent with CLEOc upper limit \cite{:2009vg} and
predictions of potential models \cite{Eichten:2002qv}.

With the same data sample, BESIII also studies \cite{Ablikim:2013gd}
$\etac(2S) \to \kskppp$ in the decay $\psp\to \gamma\etac(2S)$. 
Evidence of $\etac(2S)\to \kskppp$ is found with a statistical
significance of 4.2~$\sigma$. 
The product branching fraction is
${\cal B} (\psp\to\gamma\etac(2S)) \times {\cal B}(\etac(2S)\to \kskppp)
=(7.03\pm 2.10 ({\rm stat.}) \pm0.70 ({\rm syst.}))\times 10^{-6}$.
%%%%%%%%%%%%%%%%%%%%%%%%%%%%%%%%%%%
\section{The $\chi_{c2}(2P)$}
%%%%%%%%%%%%%%%%%%%%%%%%%%%%%%%%%%%
With a data sample of 395 fb$^{-1}$, Belle observes
\cite{Uehara:2005qd} an enhancement in the $D\bar{D}$ mass
distribution from $e^+e^- \to e^+e^- D\bar{D}$ events with a
statistical significance of 5.3 $\sigma$. 
The $D\bar{D}$ is exclusively reconstructed with four combination of
decays, 
$D^0\to K^-\pi^+$, $\bar{D}^0 \to K^+ \pi^-$; 
$D^0\to K^-\pi^+$, $\bar{D}^0 \to K^+ \pi^- \pi^0$; 
$D^0\to K^-\pi^+$, $\bar{D}^0 \to K^+ \pi^- \pi^+\pi^-$; 
$D^+\to K^-\pi^+ \pi^+$, $D^- \to K^+ \pi^- \pi^-$.
To enhance exclusive two-photon $\gamma\gamma \to D\bar{D}$
production, the total transverse momentum in the $e^+e^-$ c.m. frame
with respect to the beam direction is required to be less than
$50~\mevcc$. The mass and width are measured to be
$M=3929\pm 5({\rm stat.})\pm 2({\rm syst.})~{\rm MeV}/c^2$,
and $\Gamma =29\pm 10({\rm stat.})\pm 2 ({\rm syst.})~\mev$.
The angular distributions of candidate events are consistent with the
spin-2 helicity-2 hypothesis, and inconsistent with spin-0.
This result has been confirmed by BaBar \cite{Aubert:2010ab}.

%%%%%%%%%%%%%%%%%%%%%%%%%%%%%%%%%%%
\section{The $\psi_2(1^3D_2 ~c\bar{c})$}
\begin{figure}[htb]
\begin{center}
\includegraphics[height=2.5in]{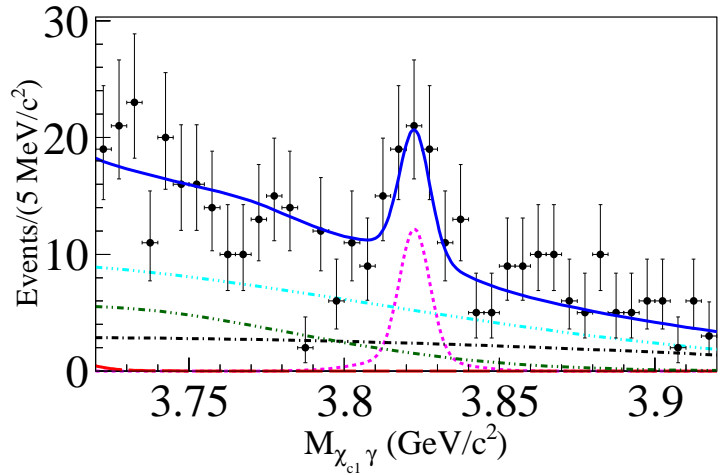}
\caption{2D UML fit projection of $M_{\chi_{c1}\gamma}$ distribution
  for the simultaneous fit of  $B^{\pm} \to (\chi_{c1} \gamma)
  K^{\pm}$ and  $B^{0} \to (\chi_{c1} \gamma) K_S^{0}$  decays for
  $M_{\rm bc} > 5.27 $ GeV$/c^2$.
The curves show the signal [red large-dashed for $\psi'$, magenta
  dashed for $X(3823)$ and violet dotted for $X(3872)$] 
and the background component 
[black dotted-dashed for combinatorial, dark green two dotted-dashed
  for $B \to\psi'(\rightarrow \chi_{cJ} \gamma) K $ and cyan three
  dotted-dashed for peaking component] as well as the overall fit
[blue solid]. $B \to\psi' (\rightarrow \chi_{cJ} \gamma)K $ is
specific to the decay mode under study.}

\end{center}
\label{fit_3823}
\end{figure}

%%%%%%%%%%%%%%%%%%%%%%%%%%%%%%%%%%
Using 772$\times 10^{6}$ $B\bar{B}$ events, Belle
observes \cite{X3823} evidence of a new resonance in the $\chi_{c1}\gamma$ final
state with a statistical significance of $3.8~\sigma$. 
The $\chi_{c1}$ is reconstructed in the decay $\chi_{c1} \to J/\psi
\gamma$, where $J/\psi$ decays to $l^+l^-$ ($l= e$ or $\mu$).
Figure \ref{fit_3823} shows the $\chi_{c1}\gamma$ invariant-mass
distribution, the solid line is the fitting result.
The mass of the state is determined to be
$3823.1\pm 1.8({\rm stat.}) \pm 0.7~({\rm syst.})~{\rm MeV}/c^2$.
The mass is near potential model expectations for the centroid of the
$1^3D_J$ states. 
No peak is seen in the $\chi_{c2}\gamma $ decay mode, the upper limit on
the ratio of the branching fraction is determined to be $R=\frac{{\cal
    B}(X(3823)\to \chi_{c2}\gamma)}{{\cal B}( X(3823)\to
  \chi_{c1}\gamma)}<0.41$ at the 90\% C.L.. This is consistent with the
expectation ($R=0.2$) for $\psi_2$ \cite{Ebert:2002pp,Ko:1997rn,Qiao:1996ve}.
The properties of the $X(3823)$ are consistent with those expected for
the $\psi_2(1^3D_2 ~c\bar{c})$ state.

%%%%%%%%%%%%%%%%%%
\section{Summary}
Charmonium is the best understood hadronic system. All the
lowest-lying charmonium states have been found in experiment. Their
properties have been measured with high precision, which are
in good agreement with theory expectation. 
Higher-mass charmonium mesons, such as $h_c(2P)$, $\chi_{cJ}'$
($J=0,1$), are still missing from experiment.
With larger data samples at different center of mass energies,
searches for the missing charmonium states will be very interesting.

%%%%%%%%%%%%%%%%%%
\Acknowledgements
I would like to thank the organizers for the successful conference. And
thank the colleagues in BESIII, Belle, BaBar and CLEOc for producing
nice results.


\begin{thebibliography}{99}

%\cite{Appelquist:1974yr}
\bibitem{1974yr} 
  T.~Appelquist, A.~De Rujula, H.~D.~Politzer and S.~L.~Glashow,
  %``Charmonium Spectroscopy,''
  Phys.\ Rev.\ Lett.\  {\bf 34}, 365 (1975).
  %%CITATION = PRLTA,34,365;%%
  %397 citations counted in INSPIRE as of 31 Oct 2013

%\cite{Godfrey:1985xj}
\bibitem{Godfrey:1985xj} 
  S.~Godfrey and N.~Isgur,
  %``Mesons in a Relativized Quark Model with Chromodynamics,''
  Phys.\ Rev.\ D {\bf 32}, 189 (1985).
  %%CITATION = PHRVA,D32,189;%%
  %1848 citations counted in INSPIRE as of 31 Oct 2013


%\cite{Siegrist:1976br} psi(4410)
\bibitem{Siegrist:1976br}
  J.~Siegrist, G.~S.~Abrams, A.~Boyarski, M.~Breidenbach, F.~Bulos, W.~Chinowsky, G.~J.~Feldman and C.~E.~Friedberg {\it et al.},
  %``Observation of a Resonance at 4.4-GeV and Additional Structure Near 4.1-GeV in e+ e- Annihilation,''
  Phys.\ Rev.\ Lett.\  {\bf 36}, 700 (1976).
  %%CITATION = PRLTA,36,700;%%
  %162 citations counted in INSPIRE as of 31 Oct 2013

%\cite{Rapidis:1977cv} psi(3370)
\bibitem{Rapidis:1977cv} 
  P.~A.~Rapidis, B.~Gobbi, D.~Luke, A.~Barbaro-Galtieri, J.~Dorfan, R.~Ely, G.~J.~Feldman and J.~M.~Feller {\it et al.},
  %``Observation of a Resonance in e+ e- Annihilation Just Above Charm Threshold,''
  Phys.\ Rev.\ Lett.\  {\bf 39}, 526 (1977)
  [Erratum-ibid.\  {\bf 39}, 974 (1977)].
  %%CITATION = PRLTA,39,526;%%
  %201 citations counted in INSPIRE as of 31 Oct 

\bibitem{Bacino:1977uh} 
  W.~Bacino, A.~Baumgarten, L.~Birkwood, C.~D.~Buchanan, R.~Burns, M.~Chronoviat, P.~E.~Condon and R.~W.~Coombes {\it et al.},
  %``Observation of a Peak in Hadron and Weak electron Production in e+ e- Annihilation at E(C.M.) = 3770-MeV,''
  Phys.\ Rev.\ Lett.\  {\bf 40}, 671 (1978).
  %%CITATION = PRLTA,40,671;%%
  %131 citations counted in INSPIRE as 
%\cite{Brandelik:1978ei}

\bibitem{Brandelik:1978ei} 
  R.~Brandelik {\it et al.}  [DASP Collaboration],
  %``Total Cross-section for Hadron Production by $e^+ e^-$ Annihilation at Center-of-mass Energies Between 3.6-{GeV} and 5.2-{GeV},''
  Phys.\ Lett.\ B {\bf 76}, 361 (1978).
  %%CITATION = PHLTA,B76,361;%%
  %111 citations counted in INSPIRE as of 31 Oct 2013

%\cite{Ablikim:2007gd}
\bibitem{BES_R_Value} 
  M.~Ablikim {\it et al.}  [BES Collaboration],
  %``Determination of the psi(3770), psi(4040), psi(4160) and psi(4415) resonance parameters,''
  eConf C {\bf 070805}, 02 (2007)
  [Phys.\ Lett.\ B {\bf 660}, 315 (2008)]
  [arXiv:0705.4500 [hep-ex]].
  %%CITATION = ARXIV:0705.4500;%%
  %60 citations counted in INSPIRE as of 31 Oct 2013


\bibitem{PDG} K.~Nakamura {\it et al.} [Particle Data Group], 
  J.\ Phys.\ G {\bf 37}, 075021 (2010).

%%%%%%%%%%%% etac paper %%%%%%%%%%%%%%
%J/psi->radiative
\bibitem{Baltrusaitis:1985mr}
  R.~M.~Baltrusaitis {\it et al.}  [Mark-III Collaboration],
  %``Hadronic Decays Of The Eta(C) (2980),''
  Phys.\ Rev.\  D {\bf 33}, 629 (1986).
  %%CITATION = PHRVA,D33,629;%%
\bibitem{Bai:2003et}
  J.~Z.~Bai {\it et al.}  [BES Collaboration],
  %``Measurements of the Mass and Full-Width of the $\eta_c$ Meson,''
  Phys.\ Lett.\  B {\bf 555}, 174 (2003).
  %  [arXiv:hep-ex/0301004].
  %%CITATION = PHLTA,B555,174;%%

%\cite{Aubert:2003pt} \\51 % two-photon etac(2S)
\bibitem{Aubert:2003pt}
  B.~Aubert {\it et al.}  [BaBar Collaboration],
  %``Measurements of the mass and width of the $\eta_c$ meson and of an
  %$\eta_c(2S)$ candidate,''
  Phys.\ Rev.\ Lett.\  {\bf 92}, 142002 (2004).
  %  [arXiv:hep-ex/0311038].
  %%CITATION = PRLTA,92,142002;%%

%\cite{Asner:2003wv}
\bibitem{Asner:2003wv}
  D.~M.~Asner {\it et al.} [ CLEO Collaboration ],
  %``Observation of eta-prime(c) production in gamma gamma fusion at CLEO,''
  Phys.\ Rev.\ Lett.\  {\bf 92}, 142001 (2004).
  % [hep-ex/0312058].

\bibitem{Uehara:2007vb}
  S.~Uehara {\it et al.}  [Belle Collaboration],
  %``Study of charmonia in four-meson final states produced in two-photon
  %collisions,''
  Eur.\ Phys.\ J.\  C {\bf 53}, 1 (2008).
  %[arXiv:0706.3955 [hep-ex]].
  %%CITATION = EPHJA,C53,1;%%

%\cite{arXiv:1105.0978}
%\bibitem{arXiv:1105.0978} 
\bibitem{2011dy}
 A.~Vinokurova {\it et al.} [Belle Collaboration],
  %``Study of B^{+-} -> K^{+-}(K_S K pi)^0 Decay and Determination of eta_c and eta_c(2S) Parameters,''  
 Phys.\ Lett.\ B\ {\bf 706}, 139 (2011).
 %[arXiv:1105.0978 [hep-ex]]. 
 %%CITATION = PHLTA,B706,139;%%

\bibitem{recent:2008fb}
  R.~E.~Mitchell {\it et al.}  [CLEO Collaboration],
  %``J/psi and psi(2S) Radiative Transitions to eta_c,''
  Phys.\ Rev.\ Lett.\  {\bf 102}, 011801 (2009).
  %[arXiv:0805.0252 [hep-ex]].
  %%CITATION = PRLTA,102,011801;%%

%**********************
%\cite{BESIII:2011ab}
\bibitem{BESIII:2011ab} 
  M.~Ablikim {\it et al.}  [BESIII Collaboration],
  %``Measurements of the mass and width of the $\eta_c$ using $\psi' -> \gamma \eta_c$,''
  Phys.\ Rev.\ Lett.\  {\bf 108}, 222002 (2012)
  [arXiv:1111.0398 [hep-ex]].
  %%CITATION = ARXIV:1111.0398;%%
  %21 citations counted in INSPIRE as of 02 Nov 2013

\bibitem{arXiv:1211.2253}
Carleton DeTar {\it et al.}, arXiv:1211.2253 (2012).

%************************************ hc
\bibitem{swanson} 
E. S. Swanson, Phys. Rep. {\bf 429}, 243 (2006).

%\cite{Dobbs:2008ec}
\bibitem{Dobbs:2008ec} 
  S.~Dobbs {\it et al.}  [CLEO Collaboration],
  %``Precision Measurement of the Mass of the h(c)(P-1(1)) State of Charmonium,''
  Phys.\ Rev.\ Lett.\  {\bf 101}, 182003 (2008)
  [arXiv:0805.4599 [hep-ex]].
  %%CITATION = ARXIV:0805.4599;%%
  %62 citations counted in INSPIRE as of 02 Nov 2013

%\cite{Ablikim:2010rc}
\bibitem{Ablikim:2010rc} 
  M.~Ablikim {\it et al.}  [BESIII Collaboration],
  %``Measurements of h_c(^1P_1) in psi' Decays,''
  Phys.\ Rev.\ Lett.\  {\bf 104}, 132002 (2010)
  [arXiv:1002.0501 [hep-ex]].
  %%CITATION = ARXIV:1002.0501;%%
  %73 citations counted in INSPIRE as of 02 Nov 2013

%\cite{Ablikim:2012ur}
\bibitem{Ablikim:2012ur} 
  M.~Ablikim {\it et al.}  [BESIII Collaboration],
  %``Study of $\psi(3686)\to\pi^0 h_c, h_c\to\gamma\eta_c$ via $\eta_c$ exclusive decays,''
  Phys.\ Rev.\ D {\bf 86}, 092009 (2012)
  [arXiv:1209.4963 [hep-ex]].
  %%CITATION = ARXIV:1209.4963;%%
  %6 citations counted in INSPIRE as of 02 Nov 2013

%\cite{Choi:2002na} belle-etac
\bibitem{Choi:2002na}
  S.~K.~Choi {\it et al.}  [Belle collaboration],
  %``Observation of the eta(c)(2S) in exclusive B ---> K K(S) K- pi+ decays,''
  Phys.\ Rev.\ Lett.\  {\bf 89}, 102001 (2002).
  %  [Erratum-ibid.\  {\bf 89}, 129901 (2002)]
  %  [arXiv:hep-ex/0206002].
  %%CITATION = PRLTA,89,102001;%%

%\cite{Nakazawa:2008zz}  %twophoton-etac(2S)
\bibitem{Nakazawa:2008zz}
  H.~Nakazawa [ Belle Collaboration ],
  %``Measurement of the eta/c and eta/c-prime mesons in two-photon process at Belle,''
  Nucl.\ Phys.\ Proc.\ Suppl.\  {\bf 184}, 220-223 (2008).
  
%\cite{Aubert:2005tj} \\double charmonium
\bibitem{Aubert:2005tj}
  B.~Aubert {\it et al.} [ BaBar Collaboration ],
  %``Measurement of double charmonium production in $e^+e^-$ annihilations at $\sqrt{s}=10.6$ GeV,''
  Phys.\ Rev.\  {\bf D72}, 031101 (2005).
  %  [hep-ex/0506062].

%\cite{Abe:2007jn}
\bibitem{Abe:2007jn}
  K.~Abe {\it et al.} [ Belle Collaboration ],
  %``Observation of a new charmonium state in double charmonium production in e+ e- annihilation at s**(1/2) ~ 10.6-GeV,''
  Phys.\ Rev.\ Lett.\  {\bf 98}, 082001 (2007).
  %[hep-ex/0507019].

%\cite{CroninHennessy:2009aa}
\bibitem{:2009vg}
  D.~Cronin-Hennessy {\it et al.}  [CLEO Collaboration],
  %``Search for psi(2S) ---> gamma eta(c) (2S) via fully reconstructed eta(c)(2S) decays,''
  Phys.\ Rev.\ D {\bf 81}, 052002 (2010)
  [arXiv:0910.1324 [hep-ex]].
  %%CITATION = ARXIV:0910.1324;%%
  %20 citations counted in INSPIRE as of 02 Nov 2013

%\cite{delAmoSanchez:2011bt}
\bibitem{delAmoSanchez:2011bt}
  P.~del Amo Sanchez {\it et al.}  [BaBar Collaboration],
  %``Observation of $\eta_c(1S)$ and $\eta_c(2S)$ decays to $K^+ K^- \pi^+ \pi^- \pi^0$ in two-photon interactions,''
  Phys.\ Rev.\ D {\bf 84}, 012004 (2011)
  [arXiv:1103.3971 [hep-ex]].
  %%CITATION = ARXIV:1103.3971;%%
  %19 citations counted in INSPIRE as of 02 Nov 2013

%\cite{Ablikim:2012sf}
\bibitem{Ablikim:2012sf} 
  M.~Ablikim {\it et al.}  [BES Collaboration],
  %``First observation of the M1 transition $\psi(3686)\to \gamma\eta_c(2S)$,''
  Phys.\ Rev.\ Lett.\  {\bf 109}, 042003 (2012)
  [arXiv:1205.5103 [hep-ex]].
  %%CITATION = ARXIV:1205.5103;%%


%\cite{Aubert:2008kp}
\bibitem{Aubert:2008kp}
  B.~Aubert {\it et al.}  [BaBar Collaboration],
  %``Study of B-meson decays to eta(c) K(*), eta(c)(2S) K(*) and eta(c) gamma K(*),''
  Phys.\ Rev.\ D {\bf 78}, 012006 (2008)
  [arXiv:0804.1208 [hep-ex]].
  %%CITATION = ARXIV:0804.1208;%%
  %23 citations counted in INSPIRE as of 02 Nov 2013

%\cite{Eichten:2002qv}
\bibitem{Eichten:2002qv}
  E.~J.~Eichten, K.~Lane and C.~Quigg,
  %``B meson gateways to missing charmonium levels,''
  Phys.\ Rev.\ Lett.\  {\bf 89}, 162002 (2002).
  %[arXiv:hep-ph/0206018].
  %%CITATION = PRLTA,89,162002;%%


%\cite{Ablikim:2013gd}
\bibitem{Ablikim:2013gd} 
  M.~Ablikim {\it et al.}  [BESIII Collaboration],
  Phys.\ Rev.\ D {\bf 87}, no. 5, 052005 (2013)
  [arXiv:1301.1476 [hep-ex]].
  %%CITATION = ARXIV:1301.1476;%%
  %1 citations counted in INSPIRE as of 02 Nov 2013

%$$$$$$$$$$$$$$$$$$$$$$$$$$$$$$$$$$$$$$$
%\cite{Uehara:2005qd}
\bibitem{Uehara:2005qd} 
  S.~Uehara {\it et al.}  [Belle Collaboration],
  %``Observation of a chi-prime(c2) candidate in gamma gamma ---> D anti-D production at BELLE,''
  Phys.\ Rev.\ Lett.\  {\bf 96}, 082003 (2006)
  [hep-ex/0512035].
  %%CITATION = HEP-EX/0512035;%%
  %198 citations counted in INSPIRE as of 31 Oct 2013

%\cite{Aubert:2010ab}
\bibitem{Aubert:2010ab} 
  B.~Aubert {\it et al.}  [BaBar Collaboration],
  %``Observation of the chi_c2(2P) meson in the reaction gamma gamma ---> D Dbar at BaBar,''
  Phys.\ Rev.\ D {\bf 81}, 092003 (2010)
  [arXiv:1002.0281 [hep-ex]].
  %%CITATION = ARXIV:1002.0281;%%
  %36 citations counted in INSPIRE as of 31 Oct 2013

%\cite{Bhardwaj:2013rmw}
\bibitem{X3823} 
  V.~Bhardwaj {\it et al.}  [Belle Collaboration],
  %``Evidence of a new narrow resonance decaying to $\chi_{c1}\gamma$ in $B \to \chi_{c1} \gamma K$,''
  Phys.\ Rev.\ Lett.\  {\bf 111}, 032001 (2013)
  [arXiv:1304.3975 [hep-ex]].
  %%CITATION = ARXIV:1304.3975;%%
  %5 citations counted in INSPIRE as of 01 Nov 2013

%\cite{Ebert:2002pp}
\bibitem{Ebert:2002pp} 
  D.~Ebert, R.~N.~Faustov and V.~O.~Galkin,
  %``Properties of heavy quarkonia and $B_c$ mesons in the relativistic quark model,''
  Phys.\ Rev.\ D {\bf 67}, 014027 (2003)
  [hep-ph/0210381].
  %%CITATION = HEP-PH/0210381;%%
  %216 citations counted in INSPIRE as of 02 Nov 2013

%\cite{Ko:1997rn}
\bibitem{Ko:1997rn} 
  P.~-w.~Ko, J.~Lee and H.~S.~Song,
  %``Color octet mechanism in the inclusive D wave charmonium productions in B decays,''
  Phys.\ Lett.\ B {\bf 395}, 107 (1997)
  [hep-ph/9701235].
  %%CITATION = HEP-PH/9701235;%%
  %26 citations counted in INSPIRE as of 02 Nov 2013

%\cite{Qiao:1996ve}
\bibitem{Qiao:1996ve} 
  C.~-f.~Qiao, F.~Yuan and K.~-T.~Chao,
  %``A Crucial test for color octet production mechanism in $Z^0$ decays,''
  Phys.\ Rev.\ D {\bf 55}, 4001 (1997)
  [hep-ph/9609284].
  %%CITATION = HEP-PH/9609284;%%
  %14 citations counted in INSPIRE as of 02 Nov 2013

\end{thebibliography}
\end{document}